# A software security review on Uganda's Mobile Money Services: Dr. Jim Spire's tweets sentiment analysis.

Authors: Nsengiyumva Wilberforce

Makerere University, College of Computing and informatics technology, Networks Department, Email: nsengiyumvawilberforce@gmail.com,

**Abstract**

The proliferation of mobile money in Uganda has been a cornerstone of financial inclusion, yet its security mechanisms remain a critical concern. This study investigates a significant public response to perceived security failures: the #StopAirtelThefty Twitter campaign of August 2025 Sparked by an incident publicized by Dr. Jim Spire Ssentongo where a phone thief accessed a victim's account, withdrew funds, and procured a loan, the campaign revealed deep seated public anxiety over the safety of mobile money. This research employs qualitative analysis to systematically examine the complaints raised during this campaign, extracting key themes related to security vulnerabilities and user dissatisfaction. By synthesizing these public sentiments, the paper provides crucial insights into the specific security gaps experienced by users and situates these findings within the larger framework of Uganda's mobile money regulatory and operational environment. The study concludes with implications for providers, policymakers, and the future of secure digital finance in Uganda.

**Source code and data**: Source code and data set used for this research

# 1 Introduction

Mobile money (MM) is a term crowned from "mobile" and "money" which when the 2 terms are combined means moving money. Mobile came from the fact that money is transferred using mobile devices such as smartphones, dumbphones, etc. Mobile money is a financial service for money transmission from one mobile device to another. In Uganda, mobile money has been widely adopted by both the banked and unbanked, from literate to the illiterate. This is due to high penetration of mobile devices within the Uganda population, it's as simple as just buying a block phone and a SIM card and you will be able to access mobile money services.

The leading providers of MM services in Uganda are Airtel Money and MTN mobile money (MoMo). With MoMo hitting 20million subscribers in 2025 ("MTN Uganda Hits 20 Million Customer Milestone," 2024) compared to Airtel money with 16.9million subscribers as of 2024 annual report(Airtel, 2024) shows that MoMo is the leading service provider in Uganda. Richard Yego, Managing Director of MTN Mobile Money Uganda Limited, highlighted that Ugandans transacted an average of UGX 159 trillion through MTN Mobile Money in 2024, equating to approximately UGX 435 billion per day (CEO East Africa, 2025)(Murungi, 2025). In 2024, Airtel Uganda demonstrated robust growth in its operations, with a total active subscriber base of 16.9 million, marking a 13.9% increase year on year. Among these, 7.31 million were data users, a 27.8% increase, while that segment's data usage per customer rose by 25.3%, contributing to a 41.7% surge in total data traffic(Airtel, 2024). This shift from traditional cash handling to electronic money transfers has significantly transformed daily financial activities. Services such as paying school fees, settling utility bills (electricity, water, waste collection), and conducting business transactions have become more efficient and secure. The widespread adoption of mobile money is evident across various sectors, including educational institutions, religious organizations, marketplaces, and public transportation services like SafeBoda. This trend underscores the growing integration of digital financial solutions into the fabric of Ugandan society.

As technology advances in the 20$^{th}$ century, security keeps on becoming a big issue because of high interconnectivity of devices both mobile and desktop devices. Security is a major component of any software service and its actual very crucial for financial services like MM. If not given bigger attention, it results into loss of monies by the subscribers. Security of MM stems from where the mobile device is kept up to the software access (PIN/password authentication and authorization) and goes also beyond to the providers (where they keep their servers, who access what access control). In August 2025, there was an X (formerly Twitter) campaign that was run by Ugandans against theft of money by people who were using Airtel Money. The campaign was called "StopAirtelTheft" and was spearheaded by a Makerere University Professor, Dr. Jim Spire Ssentongo after he lost one of his late brother's phone but could not recover money from his brother's Airtel Money account. Some of the security bleaches looked to be coming from insiders especially back office operators and others seemed to come from unauthorized PIN access, etc. We will explore more some of the issues raised by the subscribers in this paper.

In this paper, we investigate people's posts with the mentioned hash tag to analyze the security stand of MM services in Uganda.

The document is organized as follows; section 1 is Introduction where we explain what mobile money is and present its usage statistics in Uganda, section 2 consists of an overall literature review, section 3 consists of the methodology we used to do this research. Section 4 is the discussion of the results from our analysis followed by conclusion in section 5 we conclude our research with different views from our analysis.

## 2 Literature Review

Uganda's mobile money revolution began in March 2009 when MTN launched its Mobile Money service, quickly followed by operators such as Airtel and UTL. The system gained rapid adoption, with mobile money agent networks and services proliferating nationwide(Ali et al., 2020). During this formative period, the Bank of Uganda (BoU) observed the innovative shift in financial behavior and recognized the need for regulatory oversight. This culminated in BoU issuing the Mobile Money Guidelines in 2013, establishing licensing requirements, escrow mechanisms, anti money laundering provisions, and consumer protections formalizing what had already become a vital channel for financial inclusion(*Mobile Money Use: The Impact of Macroeconomic Policy and Regulation*, n.d.).

Over time, mobile money catalyzed broader financial engagement. By 2013, a FINSCOPE survey reported that 56% of adults in Uganda used mobile money, surpassing traditional bank account usage, which hovered around 20–29%(Ali et al., 2020; Katusiime, 2021). A 2016 BoU working paper applied rigorous statistical models to data from March 2009 to February 2016 and uncovered a significant long run positive relationship between mobile money uptake and private sector credit growth demonstrating how mobile money not only provided transactional convenience but also fostered formal financial intermediation. Complementing this, research by Lwanga & Adong (2016) used micro data to reveal that mobile money users were notably more likely to save particularly among urban and central region populations underscoring the platform's role in promoting household level financial discipline. By mid 2015, Uganda had over 19.8 million mobile money users about half of the country's population compared to fewer than 1,000 ATMs operated by commercial banks, illustrating how mobile money vastly outpaced traditional banking infrastructure(*Mobile Money Use: The Impact of Macroeconomic Policy and Regulation*, n.d.). In 2016, mobile money transactions reached UGX 3.6 trillion, with an average daily volume of UGX 122 billion (about USD 34 million), and the average transaction size hovered around UGX 70,000 (USD 20)(*Macroeconomic Effects of Mobile Money: Evidence from Uganda | Financial Innovation | Full Text*, n.d.). The platforms also diversified in services incorporating bill payments, remittances, savings linked products (like MoKash), micro loans, and even integration with village savings and loan associations (VSLAs)(*Home*, n.d.; "MTN Uganda," 2025).

Mobile money In Uganda keeps developing and advancing in technology as technology improves. For example, In 2025, MTN MoMo introduced a virtual card technology to enable e commerce growth (*MTN MoMo Uganda Introduces the Virtual Card by MoMo to Enable E Commerce Growth*, n.d.). MTN MoMo virtual card allows the unbanked to make online payments on any platform anywhere in the world, you just dial the USSD code (16570) and the card gets linked to your MTN MoMo account. The same year 2025, Airtel money also introduced a similar technology called Airtel Money Mastercard which is a virtual card for making online payments anywhere in the world. To acces this service on Airtel money, you need to first download Selfcare mobile app and install it on your phone, then select ACTIVATE and you will be enrolled for the service.

However, despite the goodness and the benefits of these innovations, a bigger question is security of the subscribers' money in their wallets and during transactions. Because of a bigger chain of service segments in these financial services, there is a big attack area for the attackers to gain access to the softwares. In **April 2025**, the MTN Group acknowledged a cybersecurity incident that resulted in unauthorized access to the personal information of some of its customers across various markets. While the company stated that its core financial systems remained secure, the breach raised concerns about the safety of customer data. Further details on the specific impact on Ugandan MoMo users were not immediately available(Bonisele, 2025). On the other hand, **Airtel Money in Uganda** faced a wave of public concern in **August 2025** regarding a spike in mobile money fraud. Numerous users took to social media and other platforms to report unauthorized access to their accounts. Some customers alleged that fraudsters had not only withdrawn funds but had also managed to secure loans in their names without their consent. Initially, Airtel Uganda reportedly dismissed the allegations as "misleading." However, the growing number of complaints prompted the Bank of Uganda to issue a public statement urging affected customers to utilize formal complaint channels. In a subsequent development, the Managing Director of Airtel Mobile Commerce Uganda Limited (AMCUL) issued an apology to affected customers and pledged a thorough review of the company's security protocols(Muhimba, n.d.; Odongo, n.d.).

These events unfold against a backdrop of increasing mobile money fraud across Uganda and East Africa. Broader industry reports from 2025 indicate that fraudulent activities, including SIM swap fraud, phishing scams, and unauthorized transactions, remain a significant and growing threat to the burgeoning fintech landscape in the region. The incidents involving two of the largest mobile money operators highlight the persistent challenges in securing these vital financial platforms and the critical need for continuous vigilance from both the providers and their customers.

## 3  Results Analysis

Out of 3,471 tweets scrawled from twitter, 3,357 tweets were telecom related tweets, an analysis was done on those. To arrive at this figure, we searched different keywords from our data as shown in the table 1.

| TELECOM | KEYWORDS |
|---|---|
| **MTN** | mtn, mtn mobile money, momo, mm |
| **AIRTEL** | airtel, airtel money, mm |
| **VODAFONE** | vodafone, vodacom, mm |
| **ORANGE** | orange, orange money, mm |
| **TIGO** | tigo, tigo cash, mm |
| **SAFARICOM** | safaricom, mpesa, mm |
| **OTHER** | mobile money, mobile wallet, ecocash, mm |

*Table 1Mobile money service providers and different keywords in the tweets*

The outcome shows that Airtel Money contributed a total of 2,780 tweets, MTM MoMo contributed a total of 552 tweets and the rest of the mobile money service providers contributed only 25 tweets as displayed in Figure 1. Airtel tweets are many because the data we analyzed was for the campaign that was targeted towards it. MTN mobile money also shows up because it's a direct competing mobile money service, so some complaints were comparing Airtel money with MoMo.

From the analysis, a total of 1,119 tweets were all complaints from Airtel Money and MTN mobile money with a little complaints from other mobile money service providers. We categorized the complaints as shown in the table 2

| Issue Type | Count |
|---|---|
| **fraud** | 428 |
| **network** | 249 |
| **Customer service** | 230 |
| **Service quality** | 205 |
| **charges** | 138 |
| **Pin issues** | 126 |
| **Transaction issues** | 49 |
| **Account issues** | 39 |

*Table 2Complaint categorization*

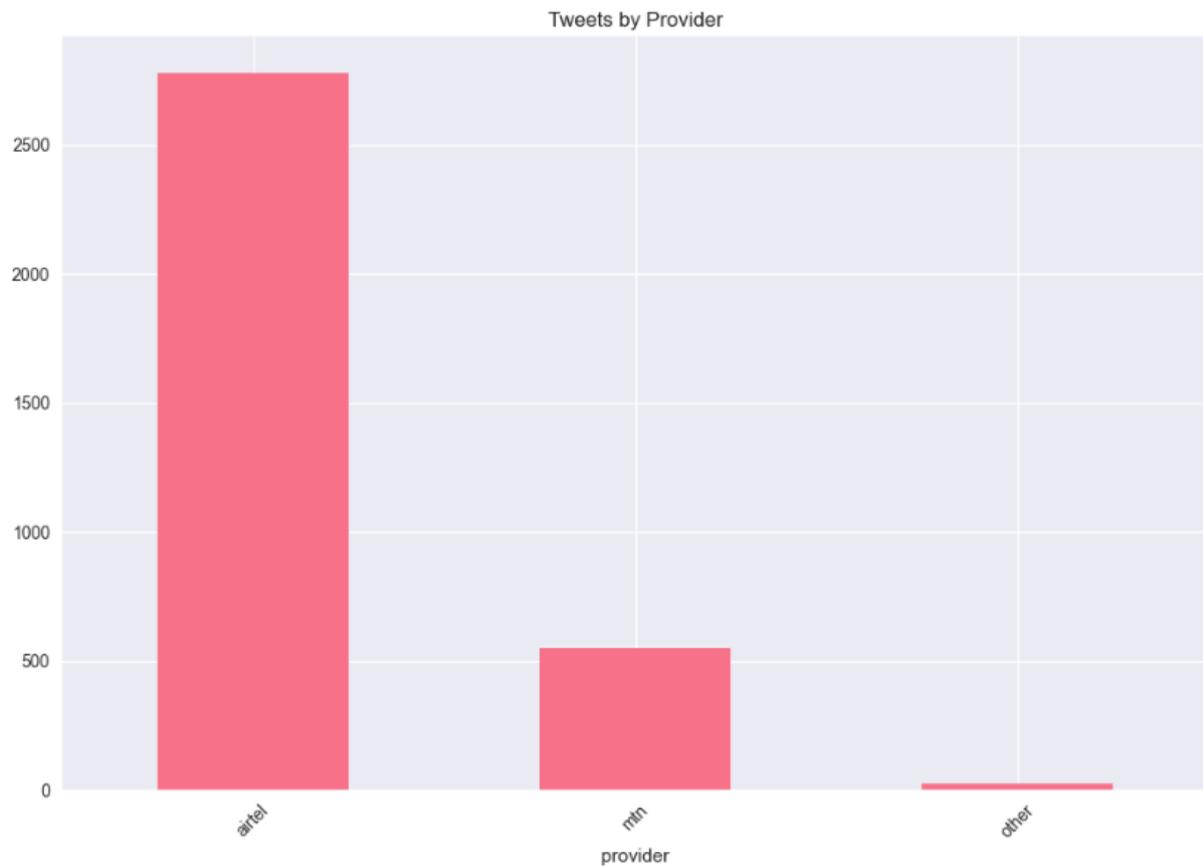

*Figure 1Tweets by mobile money service provider*

A breakdown of complaints by the network provider is shown in figure 2 and the detailed figures in table 3. Analysis of complaint data across providers reveals that **Airtel accounts for the majority of complaints (1,119 total)**, representing approximately **77% of all reported issues**. The most frequent complaints for Airtel are **fraud (30%)**, **network problems (17%)**, and **customer service issues (15%)**, indicating key areas for improvement. **MTN registers 335 complaints (23%)**, with fraud and customer service again being the leading concerns, though at lower volumes. Complaints related to **other providers are minimal (15 complaints, ~1%)**, suggesting that most customer issues are concentrated among the two main telecom companies. Overall, fraud and network issues emerge as the dominant challenges faced by users.

| Complaint Category | Airtel | MTN | Other |
|---|---|---|---|
| Account Issues | 30 | 8 | 1 |
| Charges | 105 | 30 | 2 |
| Customer Service | 171 | 58 | 1 |
| Fraud | 332 | 96 | 3 |
| Network | 191 | 54 | 4 |
| Pin Issues | 93 | 34 | 2 |
| Service Quality | 160 | 44 | 1 |
| Transaction Issues | 37 | 11 | 1 |

*Table 3A breakdown of complaint categories by Mobile money service provider*

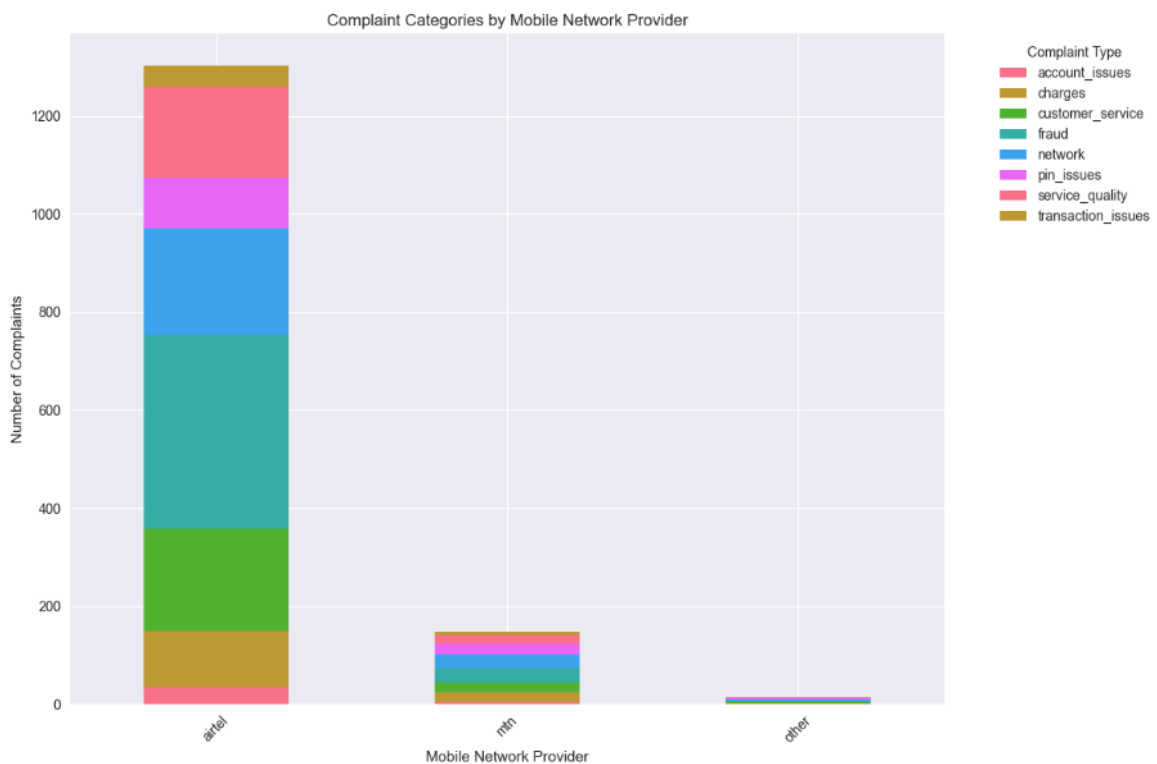

*Figure 2complaints categories by a mobile money service provider*

For the tweets that reported monetary losses, we identified approximately 125 incidents, with a total reported loss of around UGX 1,473,099,495 ($412,068). It is important to note that this figure is approximate, as our algorithm did not capture every number perfectly. Breaking down the losses by provider, Airtel accounted for UGX 778,321,970 across 97 incidents, averaging about UGX 8,023,938 per incident, while MTN reported UGX 694,777,525 over 28 incidents, with an average loss of approximately UGX 24,813,480 per incident. These

figures provide a clearer understanding of how the reported losses were distributed among the major telecom providers.

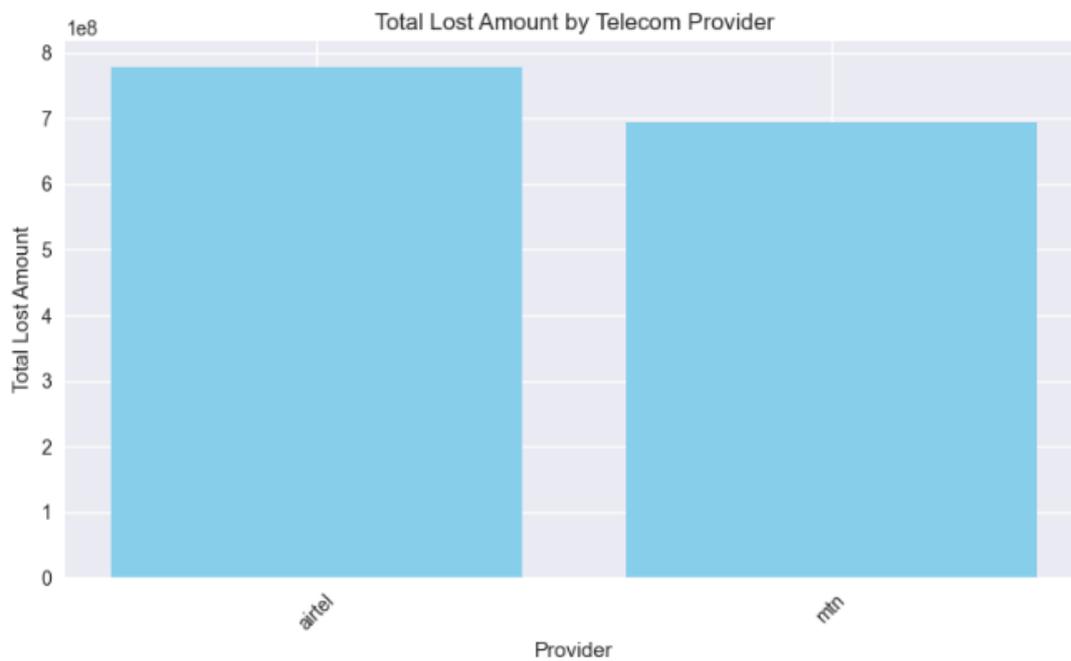

*Figure 3amount lost by each mobile money service provider*

# 4 Observations
## 4.1 Bigger overview
A bigger look the crawled tweets directed at Airtel Money Uganda, particularly through the lens of complaints that the users raised reveals widespread dissatisfaction with mobile money services, raising critical questions about service reliability, financial accountability, and customer protection within digital financial platforms in Uganda. Below are key themes that emerged from the analysis, with selected paraquoted examples for illustration.

## 4.2 Unexplained Deductions and Failed Transactions
Many users express frustration over failed transactions where funds were deducted from their Airtel Money accounts but not delivered to the intended recipients. For example, one customer lamented sending UGX 2 million to their MTN line, only for the money to vanish neither network provider resolved the issue after over two years. Similarly, another user noted losing UGX 250,000 in a failed withdrawal, with Airtel customer service repeatedly citing "network issues." In several cases, users mention losses of UGX 50,000, UGX 114,000, and UGX 803,500 under similar circumstances.

Another customer noted, "I once sent my mum UGX 3 million, and Airtel followed up the next day with threats to block her SIM unless she followed suspicious instructions. Luckily, she consulted me and ignored them." This indicates potential social engineering vulnerabilities tied to mobile money use.

### 4.3 Delayed or Absent Reversals

Several tweets suggest a systemic issue with reversals. Users routinely share that even after receiving reversal confirmation messages, the refunded funds never reflect in their accounts. One example includes a user who reported losing UGX 70,000 and receiving an SMS claiming the money had been reversed but it never appeared. Another wrote, "Sent UGX 50k from Airtel to MTN, the money was never received, customer care was no help. Stolen like that from 14 million subscribers billions gone."

### 4.4 Customer Service Gaps and Institutional Apathy

A dominant sentiment is that customer care is ineffective, dismissive, or inaccessible. One aggrieved customer stated that after calling very many times and visiting Airtel's head office, they received no help over a six month period and eventually lost over UGX 500,000. Others described being asked to present court orders to resolve issues involving amounts as low as UGX 35,000 to UGX 100,000 barriers seen as disproportionate and inaccessible for the average customer. One user wrote, "How can you tell me to go to court for my UGX 35k? On whose transport?" Another added, "I have lost so much money through customer reversals. You call, and they tell you to negotiate with the thief who reversed your transaction."

### 4.5 SIM Swap and Mobile Money Theft

Our dataset contains numerous reports of SIM card theft followed by unauthorized withdrawals and loans. In one incident, a phone was stolen, and within minutes, UGX 1.1 million was withdrawn. Airtel referred the customer to their service center but offered no resolution. In another, despite replacing a stolen SIM within hours, a user lost UGX 800,000 and incurred a UGX 100,000 loan. Victims question how such access was gained so quickly, suggesting weaknesses in PIN protection and internal security checks. One individual stated: "My uncle had UGX 4 million on his line but forgot the PIN. When he called to reset it, the money was withdrawn, and Airtel told him to go to the service center and pay to get statements."

## 4.6 Arbitrary Blocking and Account Restrictions

Several tweets describe the unexplained blocking of SIM cards and mobile money accounts after transactions involving significant amounts. One user claimed their account was locked with UGX 38 million in it, causing financial distress. Others faced similar issues even with balances as modest as UGX 3 million or UGX 4 million. "Just 4 million and you block the account? Why?" one tweet reads. In another scenario, a user who deposited UGX 1 million found the line blocked within days, with no clear reason or successful follow up. "I confronted workers at the Ishaka office, till now, I don't know how the money was withdrawn."

## 4.7 Predatory Loan Deductions and Opaque Terms

Users consistently question the fairness of Airtel's loan products. Several tweets describe taking loans of UGX 100,000 only to receive UGX 80,000 and repay the full amount with interest in a short time span. Others recount automatic deductions for quick loans they did not knowingly accept. One customer noted: "I deposited UGX 10k, and Airtel swept it saying I had a loan of UGX 5k. Where is the transparency?"

## 4.8 Loss of Confidence and Customer Migration

A growing number of users are withdrawing their funds and abandoning Airtel Money. "I read a post today, got scared, and withdrew all my UGX 5 million," said one user. Another declared, "I no longer keep more than UGX 2k on Airtel Money; even UGX 10k is too risky." Some report switching entirely to MTN or using Airtel solely for calls and data, citing better reliability elsewhere.

## 4.9 Perceptions of Systemic Theft and Lack of Accountability

Many customers accuse Airtel of systematic malpractice, frequently using terms like "thieves," "scammers," "fraudsters," and "daylight robbery." Several individuals recount losing between UGX 10 million and UGX 40 million, often with no resolution. One user stated plainly: "Airtel put me in depression after stealing my UGX 380k. It took me 3 months to recover."

There is also public suspicion that internal staff may be complicit. "I deposited UGX 1 million and received five calls from fraudsters within one week. Something fishy is happening with Airtel workers," one user noted.

# 5 Conclusion.

This study examined Uganda's Airtel Money services through the lens of the #StopAirtelTheft campaign, which brought to light widespread public concerns over the security and reliability of mobile money platforms. While mobile money has undeniably transformed financial inclusion in Uganda enabling millions to transact, save, and access credit the findings of this research point to significant and persistent vulnerabilities that undermine user trust and the sustainability of digital finance. The literature review highlighted how mobile money's rapid growth since 2009 has outpaced traditional banking infrastructure and become a pillar of Uganda's financial ecosystem. However, the same growth trajectory has also expanded the potential attack surface for fraudsters and exposed weaknesses in both technological safeguards and institutional responses. Previous studies have already signaled the systemic risks associated with SIM swap fraud, weak customer verification, and limited regulatory enforcement. Our analysis of the 2025 Airtel campaign further validates these concerns by presenting real user experiences of financial loss, failed reversals, arbitrary account blocking, and exploitative loan deductions. From the results, Airtel emerged as the provider most criticized, accounting for the majority of complaints, especially around fraud (30%), network reliability (17%), and ineffective customer service (15%). Monetary loss incidents totaled over UGX 1.4 billion, with Airtel users disproportionately affected compared to MTN. While these figures are self-reported and approximate, they underscore the scale of insecurity experienced by consumers and the tangible erosion of trust in Airtel's mobile money services. The observations reveal that security issues are not only technical but also organizational and procedural. Unexplained deductions, absent reversals, and opaque lending practices point to weaknesses in operational accountability, while reports of SIM swap fraud and insider collusion highlight deeper systemic flaws. Moreover, Airtel's slow and often dismissive customer service response has exacerbated public anger, fostering a perception of institutional indifference if not complicity. This breakdown of trust has led to visible customer migration to rival platforms, diminished confidence in digital wallets, and increased reliance on informal risk mitigation strategies, such as keeping minimal balances on mobile accounts. Taken together, the findings suggest that the crisis surrounding Airtel Money in August 2025 is not an isolated incident but a manifestation of broader structural and governance challenges in Uganda's mobile money sector. While the Bank of Uganda has established regulatory frameworks, enforcement gaps remain evident. The rapid evolution of fintech products such as virtual cards and mobile based loans requires equally adaptive oversight and stronger consumer protection mechanisms. Therefore, the implications of this research are threefold. First, for providers, there is an urgent need to strengthen technical safeguards, including multi factor authentication, advanced fraud detection, and stricter access controls to limit insider abuse. Second, for regulators, stronger enforcement of consumer protection laws and more transparent resolution mechanisms are necessary to restore public trust. Third, for users, financial literacy campaigns should accompany technological rollouts to help consumers recognize scams, understand loan terms, and demand accountability.

In conclusion, while mobile money remains indispensable to Uganda's financial inclusion journey, its future viability depends on addressing the persistent security and trust deficits that this study has brought to light. Without decisive reforms from both service providers and regulators, the promise of secure, inclusive digital finance risks being overshadowed by recurring episodes of fraud, customer dissatisfaction, and declining public confidence.

# 6 Recommendations

Based on the findings of this research, several measures are recommended to strengthen the security, accountability, and reliability of mobile money services in Uganda.

For service providers such as Airtel Money and MTN MoMo, there is an urgent need to enhance authentication mechanisms by moving beyond single-factor PIN authentication to include multi-factor authentication, biometrics, or one-time passwords in order to mitigate SIM swap fraud and unauthorized access. In addition, providers should invest in advanced fraud detection and monitoring systems capable of identifying suspicious patterns in real time, such as rapid withdrawals after SIM replacement or unusually high transaction amounts. A transparent, standardized, and auditable reversal and refund policy is equally important to ensure failed transactions are promptly resolved without forcing customers into prolonged disputes. Loan products also need greater transparency, with simplified terms communicated clearly through USSD platforms and mobile applications so that customers are fully aware of loan amounts, deductions, interest, and repayment schedules. Customer service reforms are critical and should focus on establishing responsive, well-trained teams that resolve complaints within defined timelines. Independent ombudsman models could be introduced to handle unresolved cases fairly. Furthermore, staff integrity controls must be strengthened by limiting access rights, monitoring employee activities, and instituting independent audits to detect and deter insider fraud.

For regulators such as the Bank of Uganda, the Uganda Communications Commission (UCC), and consumer protection agencies, stricter oversight and enforcement of mobile money operations are required to reflect the evolving risks in fintech. Compliance should be enforced with clear penalties against providers that fail to adequately protect customer funds. Establishing a centralized fraud reporting and tracking system would also help detect systemic patterns across providers and enable quicker regulatory intervention. A mandatory compensation policy should be introduced to require providers to compensate users in cases of negligence or unexplained transaction failures, aligning consumer protection in mobile money with formal banking practices. Regulators should also demand greater public accountability by obligating providers to publish regular reports on security incidents, fraud cases, resolutions, and systemic improvements.

Consumers themselves also have a role to play in strengthening mobile money security. Nationwide financial literacy and awareness campaigns should be launched to educate users on common fraud tactics, including phishing, social engineering, and SIM swap scams, while encouraging safe practices such as never sharing PINs. Users should also be encouraged to manage their balances wisely by limiting the amounts kept on mobile wallets to what is necessary for short-term transactions, thereby minimizing exposure to fraud-related losses. Moreover, consumers need to be empowered to make greater use of formal complaint channels provided by regulators, such as the Bank of Uganda, to ensure that grievances are documented, tracked, and addressed systematically.

Finally, cross-sector collaboration will be essential in building a more resilient digital financial ecosystem. Telecom operators, banks, and fintechs should strengthen partnerships to standardize security protocols and improve interoperability safeguards. In addition, research and innovation should be encouraged through academic–industry collaborations that continuously study emerging fraud trends and design context-specific solutions tailored to Uganda's financial environment.

Odongo, J. S. (n.d.). *Spire Welcomes Airtel's U Turn on Fraud Concerns*. Nilepost News. Retrieved August 22, 2025, from https://nilepost.co.ug/business/280870/thumbnail_22614910225557799711